\def\year{2020}\relax
\documentclass[letterpaper]{article} % DO NOT CHANGE THIS
\usepackage{aaai20}  % DO NOT CHANGE THIS
\usepackage{times}  % DO NOT CHANGE THIS
\usepackage{helvet} % DO NOT CHANGE THIS
\usepackage{courier}  % DO NOT CHANGE THIS
\usepackage[hyphens]{url}  % DO NOT CHANGE THIS
\usepackage{graphicx} % DO NOT CHANGE THIS
\usepackage{graphicx}  % DO NOT CHANGE THIS
\usepackage{subfigure}
\usepackage{threeparttable}
\usepackage{multirow}
\usepackage{amsmath}
\usepackage{algorithm}
\usepackage{algorithmic}
\usepackage{array}
\usepackage{threeparttable}
\usepackage{amsfonts}

\title{Weighted-Sampling Audio Adversarial Example Attack}
\author{Xiaolei Liu,\textsuperscript{\rm 1} Kun Wan,\textsuperscript{\rm 2} Yufei Ding,\textsuperscript{\rm 2} Xiaosong Zhang,\textsuperscript{\rm 1}\thanks{Correspondence author is Xiaosong Zhang. This research was supported by National Key R\&D Program of China (2017YFB0802900), National Natural Science Foundation of China (61572115,61902262), Sichuan Science and Technology Program (2019JDRC0069).} Qingxin Zhu\textsuperscript{\rm 1}\\ 
\textsuperscript{\rm 1}University of Electronic Science and Technology of China,   \textsuperscript{\rm 2}University of California, Santa Barbara\\ 
luxaole@gmail.com, \{kun,yufeiding\}@cs.ucsb.edu, \{johnsonzxs,qxzhu\}@uestc.edu.cn 
}
 
\begin{document}

\maketitle

\begin{abstract}
Recent studies have highlighted audio adversarial examples as a ubiquitous threat to state-of-the-art automatic speech recognition systems. Thorough studies on how to effectively generate adversarial examples are essential to prevent potential attacks. Despite many research on this, the efficiency and the robustness of existing works are not yet satisfactory. In this paper, we propose~\textit{weighted-sampling audio adversarial examples}, focusing on the numbers and the weights of distortion to reinforce the attack. Further, we apply a denoising method in the loss function to make the adversarial attack more imperceptible. Experiments show that our method is the first in the field to generate audio adversarial examples with low noise and high audio robustness at the minute time-consuming level~\footnote{We encourage you to listen to these audio adversarial examples on this website: \url{https://sites.google.com/view/audio-adversarial-examples/}.}.

\end{abstract}

\section{Introduction}

In recent years, machine learning algorithms are widely used in various fields. However, studies show that existing learning-based algorithms are vulnerable to adversarial attacks~\cite{szegedy2013intriguing,goodfellow6572explaining}. Currently, majority of the research on adversarial examples are in the image recognition field~\cite{kurakin2016adversarial,carlini2017towards,chen2017ead,su2017one}, while others investigate fields such as text classification~\cite{jia2017adversarial}, traffic classification~\cite{liu2018tltd}, and malicious software classification~\cite{grosse2016adversarial,hu2017generating,liu2019adversarial}. 

Automatic speech recognition (ASR) is another vital field where machine learning algorithms are also frequently applied~\cite{hinton2012deep}. To date, it has been proved that audio adversarial examples can mislead ASR to transfer any audio to any targeted phrases~\cite{carlini2018audio}. However, it is much more difficult to generate adversarial examples for audio than images. To generate an effective audio adversarial example, there are still several technical challenges to be addressed: 

\textbf{(C1)} Generating audio adversarial examples demands significant computational resources and huge time overhead. It takes over one hour or more to generate an effective audio adversarial example by recently proposed approaches~\cite{carlini2018audio,kreuk2018fooling,yuan2018commandersong,qin2019imperceptible}. Such inefficiency significantly undermines the practicability of the attack. 

\textbf{(C2)} Recording and replaying, which are common operations for audio, could easily introduce extra noise. Therefore, the robustness of adversarial examples against noise is crucial. Nevertheless, the adversarial examples prepared over hours are still poor in robustness.  The state-of-the-art audio adversarial examples \cite{carlini2018audio,alzantot2018did} become invalid after adding imperceptible pointwise random noise.

\textbf{(C3)} Different from the image domain where $l_p$-based metrics are carefully studied as a part of the loss function to generate adversarial examples, there are no investigations on which kind of metric is more suitable for constructing audio adversarial examples. 

In this paper, we achieve a fast, robust adversarial example attack to ASR by proposing two novel techniques named \textbf{Weighted Perturbation Technology (WPT)} and  \textbf{Sampling Perturbation Technology (SPT)}. 

WPT adjusts the weights of distortion at different positions of audio during the generation process, and thus generates adversarial examples faster and improves the attack efficiency (addressing C1). 

Meanwhile, by reducing the number of points to perturb based on the characteristics of context correlation in the speech recognition model, SPT can increase the robustness of audio adversarial examples (addressing C2). 

To best of our knowledge, we are the first in the field to both take the factors of the weights and the numbers of perturbed points into consideration during the generation of audio adversarial examples. And the two techniques are always complementary to all existing ASR adversarial attacks, which by default modify every value of the entire audio vector.

Further, we also investigate different metrics as parts of the loss function to generate audio adversarial examples and provide a reference for future researchers in this field (addressing C3). 

Finally, our experiments show that our method can generate more~\textit{robust} audio adversarial examples in a short period of~\textit{4 to 5 minutes}. This is a substantial improvement compared to the state-of-the-art methods.

\section{Related Work}
\label{relatedwork}

Audio adversarial example attacks can be mainly divided into two categories, speech-to-label, and speech-to-text~\cite{yang2018characterizing}. Speech-to-label classifies audio into different categories and the output is a specific label. This method is inspired by a similar method on images~\cite{alzantot2018did,cisse2017houdini}. Since the target phrases can only be chosen from a certain amount of labels, the practicality of such a method is limited. 

The speech-to-text method directly converts audio semantic information into text. Carlini~\& Wagner~\cite{carlini2018audio} are the first to work on audio adversarial examples for the speech-to-text models and they can let ASR transcribe any audio into a pre-specified text. However, the audio robustness is compromised and most of their examples will lose the adversarial labels by adding imperceptible random noise.

Later on CommanderSong~\cite{yuan2018commandersong} achieved practical over-the-air audio adversarial attacks, but they only validated their method on the music clips. Additionally Yakura~\&~Sakuma~\cite{yakura2018robust} proposed another physical-world attack method. Regardless, these two methods will introduce non-negligible noise to the original audio. Unfortunately, all of these methods would require several hours to generate only one audio adversarial example, including the most recent work~\cite{qin2019imperceptible}.

To the best of our knowledge, there is no method to generate audio adversarial examples with low noise and high robustness at the minute level. Our proposed method can be applied with all these current methods to achieve a trade-off among quality, robustness and convergence speed.

\section{Background}
\label{background}

\textbf{Threat Model.} Before digging into details of the audio adversarial example attack, an ASR model should be selected as the potential threat model. Following the common practice in the field we summarize three basic requirements for it:

\begin{itemize}
	\item Its core component should be Recurrent Neural Networks (RNNs) such as LSTM~\cite{hochreiter1997long}, which is widely adopted in current ASR systems;
	\item It is vulnerable to the state-of-the-art audio adversarial attack methods, and the corresponding results could be used as baselines in our experiments;
	\item It has to be open-source and thus we can directly conduct white-box tests on it.
\end{itemize}

Given requirements above, we choose the speech-to-text model, Deepspeech \cite{hannun2014deep}, as our experimental threat model, which is an open-source ASR with Connectionist Temporal Classification (CTC) method~\cite{graves2006connectionist} and LSTM as its main components. Notice that our approach can be also applied to other RNN-based ASR systems. 

Considering that there are many ways to convert the black-box model to a white-box model~\cite{papernot2016transferability,oh2017towards,ilyas2018black}, which is another research direction, and most of the previous work also assume they know the parameters of models, hence our research is also based on the white-box model.

\begin{figure}[htb]
	\centering
	\includegraphics[width=.95\columnwidth]{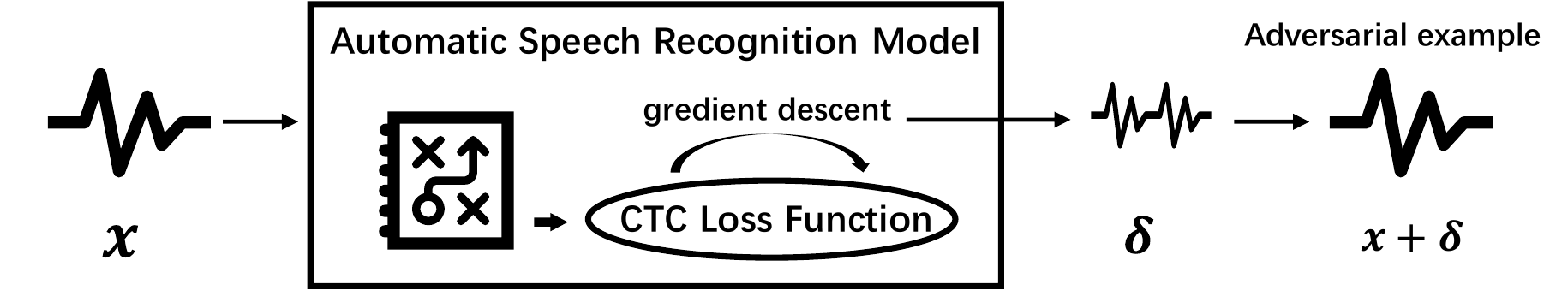}
	\caption{General process of audio adversarial example attack.}
	\label{fig:adversarial examples}
\end{figure}

\textbf{Audio Adversarial Examples.} Figure~\ref{fig:adversarial examples} shows the general process of audio adversarial example attack. Specifically, let $x$ be the input audio vector and $\delta$ is the distortion to the original audio. Audio adversarial example attacks are defined as by adding some perturbations $\delta$, ASR recognizes $x+\delta$ as specified malicious texts $t$ (formally: $f(x+\delta) = t$), while there is no perceivable difference for humans. The process of generating adversarial examples can be regarded as a process of updating $x$ using gradient descent on a predefined loss function $\ell(\cdot)$ shown in Eq.~\ref{normal-loss}. The iterative process stops until the adversarial example meets our evaluation requirements. 
\begin{equation}
\label{normal-loss}
\begin{split}
\ell(x,\delta,t) & =  \ell_{model}(f(x+\delta),t) \\
& + c \cdot \ell_{metric}(x,x+\delta)
\end{split}
\end{equation}
In Eq.~\ref{normal-loss}, $\ell_{model}$ is the loss function used in the ASR models. For example, Carlini~\& Wagner~\cite{carlini2018audio} uses CTC-loss as the $\ell_{model}$. $\ell_{metric}$ is used to measure the difference between the generated adversarial examples and the original samples. Different from the image domain where $l_p$-based metrics are commonly used, there is no consensus on which $\ell_{metric}$ should be applied in the audio field. For instance, so far various $\ell_{metric}$ such as SNR~\cite{yuan2018commandersong}, psychoacoustic hearing thresholds~\cite{schonherr2018adversarial} and frequency masking~\cite{qin2019imperceptible} have been adopted. We will also elaborate the choices of $l_{metric}$ in this paper.

\textbf{Evaluation Metric.} Based on the characteristics of the audio and the common practice in the field, the following evaluation metrics are chosen in this paper.

\begin{itemize}
	\item \textbf{SNR}(Signal-to-noise ratio) measures the noise level of the distortion $\delta$ relative to the original audio $x$. The smaller distortion is, the larger SNR will be,
	\begin{equation}
	\text{SNR} = 10\log_{10}{\frac{P_x}{P_{\delta}}},
	\end{equation}
	where $P_x$ and $P_{\delta}$ represent the energies of the original audio and the noise respectively.    
	
	\item \textbf{WER}, i.e., the word error rate, is a common evaluation metric in the ASR domain,
	\begin{equation}
	\text{WER} = \frac{S+D+I}{N} \times 100\%,
	\end{equation}
	where S, D and I are the numbers of substitutions, deletions and insertions respectively, and N is the total number of words.
	
	\item \textbf{Success Rate} is the ratio of examples which can be successfully recognized as the malicious target texts by ASR,
	\begin{equation}
	\text{Success Rate} = \frac{N_{adv}}{N_{total}} \times 100\%,
	\end{equation}
	where $N_{adv}$ is the number of adversarial examples that can be transcribed as target phrases and $N_{total}$ is the total number of adversarial examples generated.
	
	\item \textbf{Robustness Rate}. Adding noise to the audio $x$ is the same as applying transformation function $t \sim \mathcal T$ over the input $x$. Here we define the robustness rate as the success ratio of examples that can still retain adversarial property after transformed by $t(\cdot)$,
	\begin{equation}
	\label{eq:robustness}
	\text{Robustness Rate}  = \frac{N_{t(adv)}}{N_{total}} \times 100\%,
	\end{equation}
	where $N_{t(adv)}$ is the number of adversarial examples that can still be transcribed as target phrases after transformed by $t(\cdot)$.
	
\end{itemize}

\section{Methodology}
\label{methodology}

In this section, first, we will show the details of sampling perturbation technology and weighted perturbation technology. We will also explain why these methods are able to increase the robustness of adversarial examples and accelerate the attack. Finally,  we will investigate different metrics and try to find out an experiential standard to refer to, instead of directly using the $l_p$-based metrics on the image domain. 

\subsection{Sampling perturbation technology}
\label{sec:spt}

We propose SPT to increase the robustness of audio adversarial examples. It works by reducing the number of perturbed points. Here we will explain the reason why SPT works, taking the CTC loss as an example. Actually it's a general method for current audio adversarial attacks. 

We use $x$ denote an audio vector, $p$ denotes a phrase which is the semantic information of $x$ and $y$ denotes the probability distribution of $x$ decoded to $p$. $x_i$ is one frame of $x$ and $y^i$ is the probability distribution over the character which is transformed by $x_i$.

\begin{figure*}[ht]
	\centering
	\includegraphics[width=0.95\textwidth]{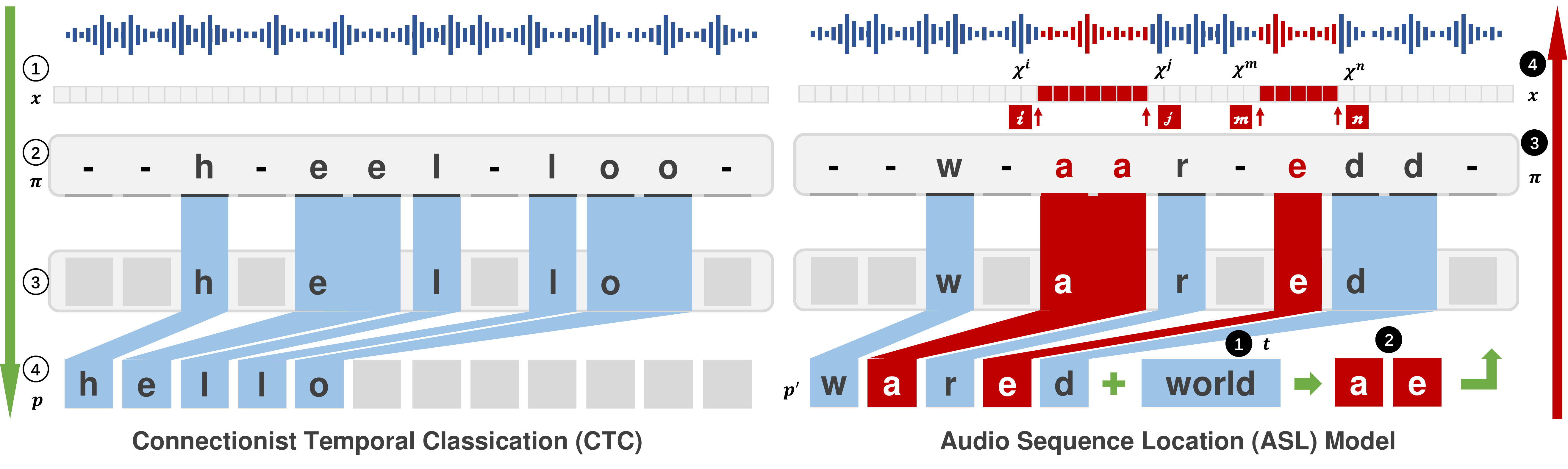}
	\caption{Overview of CTC and ASL.}
	\label{fig:ctc}
\end{figure*}

In CTC process (shown in Figure~\ref{fig:ctc} left), the process from $x$ to $p$ is: Input $x$  (Step 1) and get the sequences of tokens $\pi$ (Step 2). Then merge the repeated characters and drop `-' tokens (Step 3). Output the predicted phrase $p$ (Step 4). 

Because $\pi$ is the sequence of tokens to $x$, we say the probability of $\pi$ under $y$ is the product of the likelihoods of each $y^i_{\pi^i}$. For a given phrase $p$ with respect to $y$, there will be a set of predicted sequences $\pi \in \prod(p,y)$. Finally, we calculate $\Pr(p|y)$, the probability of phrase $p$ under the distribution $y$, by summing the probability of each $\pi$ in the set:
\begin{equation}
\label{probability_p}
\Pr(p|y) = \sum_{\pi \in \prod(p,y)} \prod_{i=0}^{n}y^i_{\pi^i}
\end{equation}
In traditional audio adversarial example attack, if we want to transcribe audio $x$ to target $t$, we will add slight distortion on each $\pi^i$ to let $t =\mathop{\arg\max}\limits_{p}\ \Pr(p|y)$. However, we can also get the same result by fixing part of $\displaystyle\prod_{j}^{n-m}y^j_{\pi^j}$ and perturbing the other part to let $\displaystyle\prod_{k}^{m}y^k_{\pi^k} = \displaystyle\prod_{k}^{m}y'^k_{\pi'^k} $, where $y'^k$ is the new probability distribution of perturbed $\pi'^k$ and $y^k_{\pi^k} \neq y'^k_{\pi'^k}$:
\begin{equation}
\label{sampling_probability_p}
\begin{split}
t  & =\mathop{\arg\max}\limits_{p}\ \Pr(p|y) \\
& =\mathop{\arg\max}\limits_{p}\  \sum_{\pi \in \prod(p,y)} \prod_{i=0}^{n}y^i_{\pi^i} \\
&=\mathop{\arg\max}\limits_{p}\  \sum_{\pi' \in \prod(p,y')} \prod_{j=0}^{n-m}y^j_{\pi^j} \prod_{k=0}^{m}y'^k_{\pi'^k} 
\end{split}
\end{equation}
Based on Formula~\ref{sampling_probability_p}, we can shorten the perturbed number of audio vector from $n$ to $m$. Our evaluations give the support that $m$ can be much smaller than $n$. 

Since most of the points in our adversarial examples are exactly the same as those in the original audio, this makes our adversarial examples show very similar properties to the original audio. Compared with the adversarial examples that all points are perturbed, environmental noise has a lower probability of affecting the SPT-based adversarial examples. 

Athalye et al.~\cite{athalye2017synthesizing} proposed the Expectation Over Transformation (EOT) algorithm to construct adversarial examples that are able to maintain the adversarial property over a chosen transformation distribution $\mathcal T$. Unfortunately, the limitation of the EOT is that it only increases robustness under the same or similar $\mathcal T$-distribution noise. Without the assumption of similar distribution, the adversarial property will be largely compromised. As a comparison, our method does not need to have prior knowledge regarding the distribution when generating adversarial examples, thus we could have better general robustness. Meanwhile, our method is complementary to EOT.

\subsection{Weighted perturbation technology}
\label{sec:wpt}

WPT can reduce the time cost by adjusting the weights of distortion in a different position. We first point out the limitations of traditional loss function $\ell(\cdot)$ (Eq.~\ref{normal-loss}) and then give our solution. (Again, we introduce WPT based on CTC sequence loss and WPT is a general method and can be easily applied to attack other ASR systems.)

\textbf{Current Problem.} By analyzing the process of generating audio adversarial examples, we found that the closer the currently transcribed phrase $p'$ is to the target text $t$, the longer it takes. In order to divide this process into different stages, we introduce the Levenshtein Distance~\cite{levenshtein1966binary}, which is a string metric for measuring the minimum number of single-character edits (i.e. insertions, deletions or substitutions) required to change one string into the other. According to our statistics, the average percentage of time loss spent on the Levenshtein distance from 3 to 2, 2 to 1 and 1 to 0 are, respectively, 7.52\%, 15.43\%, and 32.16\%. Their sum exceeds 55\% of the generation time. The reason for spending a lot of time at these stages is that when Levenshtein Distance is small, most of the current points no longer need to be perturbed, except for those points which cause the Levenshtein Distance not to be 0. We name these points as key points. 

On the one hand, if we can give these key points larger weights, the time spent at this stage will be reduced; on the other hand, if the global search step size can be reduced with the number of iterations, then we can avoid missing a more perfect adversarial example due to over perturbing. These two aspects will make the overall speed be accelerated. 

\textbf{Steps of WPT:} Accordingly, we implement WPT in two steps. The \textbf{first step} focuses on shortening the time cost when Levenshtein Distance equals to 1 by increasing the weights of key points. Therefore, we need to know which points are key points.

\textit{Audio Sequence Location(ASL)} is a model to help us locate these key points in the audio. As shown in Figure~\ref{fig:ctc} (right), the inputs of ASL are current transcribed phrase $p'$ and target $t$ (Step 1). After comparing $p'$ and $t$, we get the different characters (Step 2). Find the positions of these characters in the sequence of tokens $\pi$ (Step 3). Output the intervals set $\chi^k$ in audio vector $x$ (Step 4). Finally, the distortion corresponding to these $k$ positions in $\chi^k$ are multiplied by weights $\omega$. Our improved formulation of $\ell(\cdot)$ is,
\begin{equation}
	\label{improved_L}
	\begin{split}
		\ell(x,\delta,t) & =  \ell_{model}(f(x+\alpha \cdot \delta),t) + c \cdot \ell_{metric}(x,x+\delta), \\ \\
		\alpha_i & = 
		\left\{\begin{aligned}
		\omega ,&\ \ \  \text{if}\ \   i \in \chi^k \\
		1 , &\ \ \  \text{else}
	\end{aligned}
	\right.,\ \omega > 1,
	\end{split}
\end{equation}
where $\alpha$ is a weights vector to $\delta$, and if the vector subscript $i$ belongs to the intervals set $\chi^k$, we give these key points bigger weights $\omega$.

Besides, when we shorten Levenshtein Distance to 0, WPT goes to its \textbf{second step} to reduce the learning rate $lr$:
\begin{equation}
lr \leftarrow \beta \cdot lr,
\end{equation}
where constant $\beta$ satisfies $\beta \in (0,1)$. After updating $lr$, we can calculate the perturbations $\delta$ on each iteration:
\begin{equation}
\delta_0 = 0,\ \ \delta_{n+1} \leftarrow \delta_{n}  - lr \cdot sign (\nabla_{\delta}\ell(x,\delta,t)),
\end{equation}
where $\nabla_{\delta}\ell(x,\delta,t)$ is the gradient of $\ell$ with respect to $\delta$.

\textbf{Advantages:} Carlini\&Wagner try to set different weights to each character of the sequence of $\pi$ to solve this problem \cite{carlini2018audio}. 
Actually it will cost prohibitive computation to find the most suitable weight for each character. So, they have to get a feasible solution $x_0$ which is found by using the normal CTC-loss function first and then using their improved method based on $x_0$. However, this is not a perfect solution to solve the problem mentioned before. There are three advantages to our WPT:

1. Their method has to find a feasible solution $x_0$ first, which means they can not shorten the time cost before generating a successful adversarial example. This period time accounts for more than 55\% of the total time. We can use ASL at any iterations to get the key location intervals $\chi^k$ without having to obtain $x_0$ first. Then we make converge faster by adjusting the weight $\omega$ of $\delta$.

2. WPT is effective against both a greedy decoder and beam-search decoder~\cite{graves2006connectionist}, which are two searching ways combined with CTC to obtain the alignment $\pi$, while their method is only effective against greedy decoder. The reasons are \textbf{a)} Instead of adjusting the weight of a single character or token, we adjust the weights of a continuous interval on the audio vector corresponding to the character. This distortion based on the continuous interval is effective for beam-search decoder. \textbf{b)} WPT updates weights $\omega$ according to the current alignment $\pi$ instead of a fixed $\pi_{0}$. So our method won't be limited to the greedy decoder.

3. The learning rate $lr$, that gradually decreases as the distortion $\delta$ is reduced, can help us avoid the problem of excessive perturbations due to too long steps so that better adversarial examples can be found more quickly.

\begin{table*}[htb]
	\centering
	\begin{threeparttable}
		\caption{Evaluation of our adversarial attack with Commander Song and C\&W's attack.}
		
		\label{tab:exp2}    
		\begin{tabular}{cp{2.5cm}<{\raggedright}p{1.8cm}<{\centering}p{2.1cm}<{\centering}p{2.1cm}<{\centering}cp{1cm}<{\centering}}
			\hline
			\hline
			Attack Approach   & Target phrase & Proportion $\downarrow$  & Efficiency(s) $\downarrow$ & Success Rate $\uparrow$ & $dB_x(\delta)$ $\uparrow$ \tnote{**}& SNR $\uparrow$\\ \hline
			{Our attack} & Random phrases\tnote{*}   & \textbf{75\%}     & \textbf{251}      & \textbf{1}       & \textbf{46.92}  & \textbf{31.9}  \\   \hline              
			
			{C\&W's attack} &  Random phrases\tnote{*}   & all points     & $\approx$3600      & \textbf{1}       & 38  & -\tnote{**}  \\   \hline  
			
			\multirow{2}{*}{CommanderSong}      & echo open the front door       & all points          & 3600      & \textbf{1}  & - \tnote{**}   & 17.2     \\ 
			& okay google restart phone now       & all points          & 4680                  & \textbf{1}    & -\tnote{**}      & 18.6           \\ \hline
			\hline                             
			
		\end{tabular}
		\begin{tablenotes}
			\footnotesize
			\item[*] As is selected in C\&W's work: target phrase is chosen at random such that (a) the transcription is incorrect (b) it is theoretically possible to reach that target. 
			\item[**] $dB_x(\delta)$ is a $l_{\infty}$ metric defined by C\&W~\cite{carlini2018audio}. And`-' means no relevant data was provided in their papers. `$\uparrow$' means the bigger the better.
		\end{tablenotes}
	\end{threeparttable}
\end{table*}

\subsection{Investigation of metrics}

As for $\ell_{metric}$, which is the other part of $\ell(\cdot)$, also plays an important role in the generation of adversarial audio. Different from the image domain where mainly $l_p$-based metrics are used as $\ell_{metric}$, there is no study on which metric should be selected. 

The purpose of $\ell_{metric}$ is to limit the difference between the adversarial examples and the original samples. Therefore, we introduce the Total Variation Denoising (TVD) to reduce the noise perturbed and let adversarial examples sound more like the original audio. TVD is based on the principle that signals with excessive and possibly spurious detail have high total variation and is mostly used in the process of noise removal \cite{rudin1992nonlinear}. After the TVD process, we can remove most of the impulse in the adversarial examples and make the distortion more imperceptible. The $\ell_{metric}$ based on TVD can be calculated via the sum of closeness $E(\delta)$ and total variation $V(x+\delta)$:
\begin{equation}
	\label{tvd}
	\begin{split}
		\ell_{metric}^{tvd}(x,\delta) & = E(\delta) + \gamma \cdot V(x+\delta) \\
		& = \frac{1}{n} \sum_{i=0}^n(\delta_i)^2 + \gamma \cdot \sum_{j=1}^{n-1} |(x_{j+1} \\ 
		& +\delta_{j+1})-(x_j+\delta_j)| ,
	\end{split}
\end{equation}
where $\gamma$ is a trade off adjusted of $E(\delta)$ and $V(x+\delta)$. Besides, we also investigate other three types of similarity metrics which are selected in terms of 1) $l_{\infty}$ in image domain; 2) $l_2$-based in current audio domain; and 3) cosine distance in information retrieval domain; as shown in Formula~\ref{Linput}.
\begin{align}
	\label{Linput}
	\begin{split}
		\ell^1_{metric}(x, \delta) & = l_{\infty}(x, x+\delta) \\
		\ell^2_{metric}(x, \delta) & = l_2(x, x+\delta)  \\
		\ell^3_{metric}(x, \delta) & = (1-cor(x, x+\delta)) 
	\end{split},
\end{align}
where $l_{\infty}(\cdot)$, $l_2(\cdot)$ and $cor(\cdot)$ are, respectively, the measurement of $l_{\infty}$ distance, $l_2$ distance and cosine distance between two audio vectors. 

A good choice of $\ell_{metric}$ not only accurately reflects the auditory difference between the two audio frequencies but also avoids the optimization process oscillating around a solution without converging \cite{carlini2017towards}. We will give a comparison of the effects of various loss functions in the experimental section.

\section{Experimental results}
\label{experiments}

In this section, we show the evaluation of our adversarial attack used the technologies introduced in the Methodology Section. We also study the performance of different $\ell_{metric}$ on success rate, SNR and $dB_x(\delta)$. Our experimental results show that our approach has faster generation speed, better SNR, higher success rate, and stronger robustness than other attacks. 

\subsection{Dataset and experimental settings}
\textbf{Dataset.} Mozilla Common Voice dataset\footnote{\url{https://voice.mozilla.org/en/datasets}} (MCVD): MCVD is an open and publicly available dataset of voices that everyone can use to train speech-enabled applications. It consists of voice samples require at least 70GB of free disk space. We follow the convention in the field and use the first 100 test instances of this dataset to generate audio adversarial examples. \textbf{Unless otherwise specified, all our experimental results are averaged over these 100 instances.}

\textbf{Environment.} All experiments are carried out on an Ubuntu Server (16.04.1) with an Intel(R) Xeon(R) CPU E5-2603 @ 1.70GHz, 16G Memory and GTX 1080 Ti GPU.

\subsection{Experiments}
\subsubsection{Evaluating adversarial examples}
In order to illustrate the effectiveness of our approach, we compared it with other two methods, Carlini \& Wagner's attack \cite{carlini2018audio} and CommanderSong \cite{yuan2018commandersong}. Table~\ref{tab:exp2} gives the success probability, average SNR, $dB_x{(\delta)}$ and efficiency for our method and two other state-of-the-art methods. 
For our method, we use SPT and WPT to improve the generation, use Eq.~\ref{sampling_probability_p} as $\ell_{model}$ and set $\ell_{metric}^{tvd}$ as our $\ell_{metric}$ (Eq.~\ref{tvd}), and the proportion of perturbed points is chosen to be 75\%.

As shown in Table~\ref{tab:exp2}, our fast approach shortens the generation time from one hour to less than 5 minutes by focusing on the key points and dynamic learning rate to accelerate the converge. In addition, our adversarial examples also have a better average of $dB_x(\delta)$ and SNR, that is, we use less calculation time and get better results. More importantly, our approach has better robustness which is shown in the next section.

\begin{table*}[htb]
	\centering
	\caption{The robustness against noise from $\bigtriangleup = 5$ to $\bigtriangleup =30$.}
	\begin{tabular}{c|cc|cc|cc}
		\hline 
		\hline 
		\multirow{2}[4]{*}{Approach} & \multicolumn{2}{c|}{$\bigtriangleup=5$} & \multicolumn{2}{c|}{$\bigtriangleup=15$} & \multicolumn{2}{c}{$\bigtriangleup=30$}  \\
		& Robustness $\uparrow$ & WER $\downarrow$  & Robustness $\uparrow$ & WER $\downarrow$  & Robustness $\uparrow$ & WER $\downarrow$ \\
		\hline 
		baseline (C\&W's attack)     & 0.23 &0.49 & 0.04  & 0.81 & 0.01 & 0.93 \\
		EOT-based ($\bigtriangleup=5$)     & 0.25    & 0.46 &0.06  &0.74  & 0.02  & 0.94 \\
		EOT-based ($\bigtriangleup=15$)    & 0.63   &0.16  & 0.07   & 0.56   &0.02   & 0.92  \\
		EOT-based ($\bigtriangleup=30$)    & 0.74  & 0.04    &0.29    & 0.23  & 0.04  & 0.70  \\
		SPT-based (proportion =5\%)     & 0.71   &0.04   & 0.4   &0.22   &0.27   &0.39   \\
		SPT-based (proportion =30\%)    & 0.58   &0.15    & 0.21   & 0.33   & 0.11    & 0.58   \\
		SPT-based (proportion =75\%)    & 0.42  &0.21   & 0.18   &0.50 & 0.06  &0.72 \\
		SPT-EOT-based (75\%,30)  & 0.85  & 0.03  & 0.30  & 0.19  & 0.09  & 0.53 \\
		\hline 
		\hline 
	\end{tabular}%
	\label{tab:robustness}%
\end{table*}

\begin{table*}[htb]
	\centering
	\begin{threeparttable}
		\caption{Evaluation of different loss functions in our adversarial attacks.}
		\begin{tabular}{p{5cm}<{\centering}p{2cm}<{\centering}p{2cm}<{\centering}p{3cm}<{\centering}}
			\hline \hline
			Loss functions \tnote{*} & SNR $\uparrow$  & $dB_x(\delta)$ $\uparrow$  & Success Rate $\uparrow$ \\
			\hline
			$\ell_0 = \ell_{model} + c_0 \cdot \ell_{metric}^{tvd}$    & \textbf{31.9}  & \textbf{46.92} & \textbf{1} \\
			$\ell_1 = \ell_{model} + c_1 \cdot \ell_{metric}^{1}$    & 29.17 & 44.55 & 0.97 \\
			$\ell_2 = \ell_{model} + c_2 \cdot \ell_{metric}^{2}$    & 30.2  & 44.91 & 1 \\
			$\ell_3 = \ell_{model} + c_3 \cdot \ell_{metric}^{3}$    & 30.1  & 44.63 & 0.98 \\
			\hline \hline
		\end{tabular}%
		\label{tab:exp1}
		\begin{tablenotes}
			\footnotesize
			\item[*] We tried our best to tune every constant $c$ of different $\ell_{metric}$ for a fair comparison. We refer interested readers to Implementation Details Section for setting details.
		\end{tablenotes}
	\end{threeparttable}
\end{table*}

\subsubsection{Evaluating robustness to noise} 
\label{exp:robustness}

As we mentioned in Eq.\ref{eq:robustness}, we evaluate the robustness of audio adversarial examples by adding noise to them and checking their adversarial properties. The process of adding noise is equal to apply transformation function $t \sim \mathcal T$ over the input audio. In our experiments, we set $\mathcal T$ as the uniform distribution with the boundary of $\pm \bigtriangleup$. We respectively added noise to SPT-based, EOT-based and SPT-EOT-based adversarial examples. Then we transcribed the newly obtained audio and finally calculate WER and Robustness Rate. If the newly transcribed phrase is the same as before, we say that this audio successfully bypassed the noise defense. 

As shown in Table~\ref{tab:robustness}, mostly the SPT-based method performs better than the EOT-based method in terms of WER. The EOT-based audio has a higher Robustness Rate when its distribution is the same or similar to the noise distribution. However, the SPT-based audio exhibits more general robustness. More specifically, in SPT, the smaller the proportion, the better the robustness, but too small proportion results in a decrease in SNR and success rate. Fortunately, the SPT-EOT-based approach combines the advantages of both methods and performs well in all aspects. We recommend using the SPT-EOT-based approach to increase robustness in future work.

\subsubsection{\textbf{Investigation of different $\ell_{metric}$}}
\label{lossfunction}
In this experiment, we generate adversarial examples based on $\ell_{model}$ and four different $\ell_{metric}$ (from Eq.~\ref{tvd} to Eq.~\ref{Linput}). For each specific loss function, we conduct adversarial attacks both with SPT (under the proportion of 75\%) and WPT.

The results in Table-\ref{tab:exp1} suggest that $\ell_0$ has the best performance on SNR,  $dB_x(\theta)$ and success rate. Besides, because the TVD process eliminates the harsher impulse noise, the added perturbation "sounds" more imperceptible. As a result, although the SNR and $dB_x(\theta)$ of $\ell_0$ are not greatly improved in numerical value, its adversarial audio sounds quite better. Here we again suggest you listen to our demos on the website has given before.

The overall performances of $\ell_1$ and $\ell_3$ are not satisfactory. Since the maximum value in the audio vector is impossible to measure the magnitude of the two small disturbances under the same maximum value. It also proves that the character of the cosine distance is more suitable for audio similarity measurement. Because $l_2$ distance can reflect all the perturbation of audio, $\ell_2$ has a better performance than $\ell_1$ and $\ell_3$, especially on the success rate. However, it's still worse than $\ell_0$ on SNR and  $dB_x(\theta)$.

Combined the experimental results and the process of tuning, we conclude that a good loss function should satisfy the following three characteristics: 1) The value ranges of $\ell_{model}$ and $c \cdot \ell_{metric}$ should be the same order of magnitude; 2) It should ensure that the value of $\ell_{model}$ are relatively larger in the initial stage, so that a feasible solution can be found as soon as possible; after finding a feasible solution, the weight of $\ell{metric}$ should increase, because it is necessary to find a feasible solution that is closer to the original sample; 3) Considering the characteristic of sound, a metric in audio area instead of general metric can lead to a more imperceptible adversarial audio.

\subsection{Implementation Details}

For reproducibility, here we give the hyperparameters used in our experiments. 

\subsubsection{Evaluating adversarial examples}
\label{appendix:exp1}

In this experiment, we generate audio adversarial examples with SPT and WPT and we select $\ell_{metric}^{tvd}$ as $\ell_{metric}$. The searching decoder is a beam-search decoder. The max iteration is set to be 500, which is enough for our method to generate imperceptible adversarial examples. In SPT, the proportion of perturbed points is 75\%. In WPT, we set the weights of key points to be 1.2, the learning rate begins with 100 and $\beta$ is set to be 0.8. $lr$ will be updated by $\beta \cdot lr$, if the $\text{iterations} \% 50 == 0$ and we have generated at least one adversarial example by now. The hyperparameters $c$ and $\gamma$ are 0.001 and 10.

\subsubsection{Evaluating robustness to noise}

Most of the hyperparameters are set to be the same as the first experiment except that the proportion of perturbed points are 5\%, 15\%, 30\%, 75\%, respectively.

\subsubsection{Investigation of different $l_{metric}$}

Most of hyperparameters are set to be the same as the first experiment. And $c_1$, $c_2$, $c_3$ are 0.01, 0.001, 1, respectively.

\subsection{Transcription Examples}

Some of the transcription examples are shown in Table~\ref{transcription examples}. All of the phrases are selected randomly from the MCVD.

\begin{table}[htb]
	\begin{center}
		\caption{Some of the transcription examples. }
		\label{transcription examples}
		\begin{tabular}{c|p{4.9cm}}
			\hline \hline
			Original phrase 1& he thought of all the married shepherds he had known \\
			\hline
			Targeted phrase 1& we're refugees from the tribal wars and we need money the other figure said \\
			\hline \hline
			Original phrase 2& i told him we could teach her to ignore people who waste her time \\
			\hline
			Targeted phrase 2& down below in the darkness were hundreds of people sleeping in peace \\
			\hline \hline
			Original phrase 3& but finally the merchant appeared and asked the boy to shear four sheep \\
			\hline
			Targeted phrase 3& it seemed so safe and tranquil \\
			\hline \hline
			Original phrase 4 & this is no place for you  \\
			\hline
			Targeted phrase 4& but finally the merchant appeared and asked the boy to shear four sheep \\
			\hline \hline
			Original phrase 5& some of the grey ash was falling off the circular edge \\
			\hline
			Targeted phrase 5& we're refugees from the tribal wars and we need money the other figure said \\
			\hline \hline
		\end{tabular}
	\end{center}
\end{table}

\subsection{Notations and Definitions}

All notations and definitions used in our paper are listed in Table~\ref{notations}.

\begin{table}[htb]
	\begin{center}
		\caption{Notations and Definitions used in our paper. }
		\label{notations}
		\begin{tabular}{c|p{6cm}}
			\hline \hline
			$x$& The original input audio  \\
			\hline
			$\delta$& The distortion to the original audio \\
			\hline
			$t$ & The targeted texts \\
			\hline
			$f(\cdot)$ & The threat model \\
			\hline
			$\ell(\cdot)$ & The loss function to generate audio adversarial examples \\
			\hline
			$\ell_{model}(\cdot)$& The loss function to measure the difference between the current output of the model and the targeted texts \\
			\hline
			$\ell_{metric}(\cdot)$& The loss function to limit the difference between the adversarial examples and the original samples \\
			\hline
			$p$& The phrase of the semantic information of original audio \\
			\hline
			$p'$& The current transcribed phrase by ASL \\
			\hline
			$y$& The probability distribution over the transformed characters \\
			\hline
			$\pi$& The sequence of tokens \\
			\hline
			$n$ & The length of the original audio vector  \\
			\hline
			$m$& The length of the perturbed audio vector \\
			\hline
			$\chi$ & the key location interval set \\
			\hline
			$c$ & A hyperparameter to balance the importance of $\ell_{model}$ and $\ell_{metric}$\\
			\hline
			$\omega$& The weights of key points \\
			\hline
			$\alpha$ & The weights of $\delta$ \\
			\hline
			$lr$ & The learning rate in gradient descent \\
			\hline
			$\beta$ & A hyperparameter to control the rate of decrease of the learning rate\\
			\hline
			$\nabla_{\delta}\ell(\cdot)$ & The gradient of $\ell(\cdot)$ with regard to $\delta$\\
			\hline
			$E(\cdot)$ & The sum of closeness \\
			\hline
			$V(\cdot)$ & The total variation \\
			\hline
			$\gamma$ &  A hyperparameter to balance the importance of $E(\cdot)$ and $V(\cdot)$ \\
			\hline
			$l_p(\cdot)$ & The $l_p$ distance, such as $l_0$, $l_2$, and $l_{\infty}$ etc. \\
			\hline
			$cor(\cdot)$ & The cosine distance \\
			\hline \hline
		\end{tabular}
	\end{center}
\end{table}

\section{Conclusion}
\label{conclusion}
This paper proposes a weighted-sampling audio adversarial example attack. The experimental results show that our method has faster speed, less noise, and stronger robustness. More importantly, we are the first to introduce the factor of the numbers and weights of perturbed points into the generation of audio adversarial examples. We also introduce TVD to improve the loss function. The study of the effectiveness of loss function shows there are some differences between the adversarial examples of image and audio. It also guides us on how to construct a more appropriate loss function in the future. Our future work will focus on the defense of audio adversarial examples.

\bibliographystyle{aaai}
\bibliography{final-version}

\end{document}